\documentclass[%
 reprint,
superscriptaddress,
 amsmath,amssymb,
 aps,
]{revtex4-1}

\usepackage{graphicx}
\usepackage{dcolumn}
\usepackage{bm}
\usepackage{natbib}
\usepackage[export]{adjustbox}
\usepackage{tikz}
\usepackage{appendix}
\usepackage{xcolor}
\usepackage{float}
\graphicspath{Figures}

\usepackage{xr}
\makeatletter
\newcommand*{\addFileDependency}[1]{
  \typeout{(#1)}
  \@addtofilelist{#1}
  \IfFileExists{#1}{}{\typeout{No file #1.}}
}
\makeatother

\usepackage{natbib}
\usepackage[export]{adjustbox}
\usepackage{tikz}
\graphicspath{Figures}

\begin{document}


\title{Printable, castable, nanocrystalline cellulose-epoxy composites exhibiting hierarchical nacre-like toughening}

\author{Abhinav Rao}
\affiliation{Department of Mechanical Engineering, Massachusetts Institute of Technology, Cambridge MA 02139, USA
}
 
\author{Thibaut Divoux}
\affiliation{Centre de Recherche Paul Pascal, CNRS UMR 5031 - 115 avenue Schweitzer, 33600 Pessac, France}
\affiliation{MultiScale Material Science for Energy and Environment, UMI 3466
CNRS-MIT,\\ 77 Massachusetts Avenue, Cambridge, Massachusetts 02139, USA}

\author{Crystal E. Owens}
\affiliation{Department of Mechanical Engineering, Massachusetts Institute of Technology, Cambridge MA 02139, USA
}

\author{A. John Hart}
\email{Corresponding author: ajhart@mit.edu}
\affiliation{Department of Mechanical Engineering, Massachusetts Institute of Technology, Cambridge MA 02139, USA
}


\begin{abstract}
Due to their exceptional mechanical and chemical properties and their natural abundance, cellulose nanocrystals (CNCs) are promising building blocks of sustainable polymer composites. However, the rapid gelation of CNC dispersions has generally limited CNC-based composites to low CNC fractions, in which polymer remains the dominant phase. Here we report on the formulation and processing of crosslinked CNC-epoxy composites with a CNC fraction exceeding 50 wt.\%. The microstructure comprises sub-micrometer aggregates of CNCs crosslinked to polymer, which are analogous to the lamellar structure of nacre and promotes toughening mechanisms associated with bulk ductile behavior, despite the brittle behavior of the aggregates at the nanoscale. At 63 wt.\% CNCs, the composites exhibit a hardness of 0.66 GPa and a fracture toughness of 5.2 MPa.m$^{1/2}$. The hardness of this all-organic material is comparable to aluminum alloys, and the fracture toughness at the centimeter scale is comparable to that of wood cell wall. We show that CNC-epoxy composite objects can be shaped from the gel precursors by direct-write printing and by casting, while the cured composites can be machined into complex 3D shapes. The formulation, processing route, and the insights on toughening mechanisms gained from our multiscale approach can be applied broadly to highly loaded nanocomposites.
\end{abstract}

\maketitle


\section{Introduction}
\label{intro}
Cellulose is nature's most abundant polymer, and an estimated 10$^{10}$ to 10$^{11}$ tons of cellulose are synthesized and naturally degraded each year \cite{Hon1994}. Cellulose nanocrystals (CNCs), which can be extracted from natural cellulose fibers by acid hydrolysis  \cite{Lagerwall2014,Habibi2010}, are of growing interest as ingredients in the production of sustainable composites due to their attractive mechanical properties, including an axial Young's modulus of 150~GPa, and their sustainability and biocompatibility \cite{Moon2011,Hon1994,Kim2015}. 

Processing CNC composites with uniform dispersion and strong bonding between CNCs and polymers remains challenging because suspensions of CNCs are susceptible to gelation and phase separation at relatively low concentrations \cite{Klemm2011,George2015}. Previous studies have shown that CNC aggregates reinforce polymer films resulting in improved mechanical properties and thermal stability \cite{Khelifa2016,Pruksawan2020TougheningAdhesive,Siqueira2017}. Yet, there remains an opportunity for further analysis of the multiscale relationships between processing, microstructure, and mechanical properties of CNC-polymer composites.

More broadly, naturally occurring nanocomposites provide compelling inspiration for the design of new synthetic materials. For instance, the high specific strength and toughness of wood originate from the architecture of its cell walls \cite{Wegst2015,Barthelat2016,Adler2013}, which comprise parallel filaments of cellulose linked by a matrix of hemicellulose and lignin arranged in layers with micro-scale thickness. 
Mineral-based nanocomposites such as nacre, enamel, and bone achieve high strength and toughness, via optimized microstructures that enable hierarchical damage resistance \cite{Ritchie2011,Dunlop2010,Dunlop2011,Pro2019,Wegst2015}, such as the brick and mortar microstructure of nacre \cite{Gu2017}.  Therein, stiff, brittle platelets are arranged in a lamellar structure, linked together with a small volume fraction of a relatively soft polymeric interfacial material. Nacre-inspired composites have been synthesized using a variety of platelet materials including graphene, ceramics, and rigid polymers, with dimensions spanning from nanometers to millimeters in length \cite{Li2012a,Zhao2018,LeFerrand2015,Gao2017,Wang2009,Gu2017}. In general, anisotropic interfaces in these composites promote fracture toughening mechanisms such as crack deflection, bridging and branching, that result in a high fracture toughness without significantly compromising the stiffness and strength provided by the individual platelets. The nacre-like microstructure can confer a ductile behavior to conventionally brittle materials, such as ceramics \cite{Bouville2014}. 

Here, we report on the formulation, mechanical properties, and near-net-shape fabrication of strong and tough CNC-epoxy composites with CNC loading exceeding 50\% wt. A gel precursor comprising CNCs and epoxide oligomers dispersed in a solvent, is printed by direct ink writing and crosslinked to form dry, solid CNC-epoxy nanocomposites. This gel-based approach allows viscous flow necessary for printing and near-netshape molding, and also a high CNC loading in the composite. The resulting composites are characterized mechanically using a combination of atomic force microscopy (AFM), statistical nanoindentation, microindentation, and scratch tests. The composites display a nanoscale granular structure resembling the brick and mortar architecture of nacre, and the grain size plays a key role in determining the mechanical properties of the composite. Notably, the micro-mechanical response of the composite resembles that of a ductile material, despite an inherently brittle behavior at the nanoscale. While similar effects have been observed for composites with inorganic fillers, we use multiscale mechanical testing to demonstrate the influence of CNC chemistry and processing in determining the behavior of the composite. At the microscale, we observe that the toughening is a consequence of fracture toughening mechanisms such as crack branching, bridging, and deflection.

\begin{figure*}
    \centering
	\includegraphics[width=1.3\columnwidth]{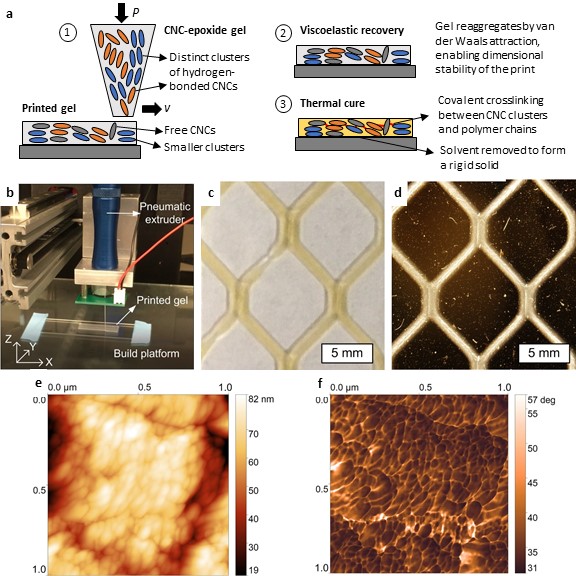}
	\caption{Fabrication of CNC-epoxy composites: a) Schematic showing the microstructure of the gel precursor during printing and post-cure processing sequence. The gel is initially composed of locally aligned CNC domains, within which the CNCs are linked by hydrogen bonds. These hydrogen bonds are partially broken during shear flow imposed by extrusion. After printing, the gel is dried and thermally cured to produce the final, dense composite. b) The 3D printer with a numerically controlled pneumatic extruder and a heated print bed. c) Bright light and d) Cross-polarized optical microscope image of the CNC-epoxide composite printed in a hexagonal lattice pattern showing bright nematic domains and good shape retention of the traces e,f) AFM topography and phase images (1~$\mu$m$^2$ area), respectively, reveal that the printed composites have a granular microstructure, with grain size of around 100~nm.}
	\label{Fig01}
\end{figure*}

\section{Results and Discussion}
\label{sec:1}
The CNC-epoxy composites are formulated from gel precursors composed of CNCs, epoxide oligomers (bisphenol A-co-epichlorohydrin glycidyl end-capped), and a solvent (dimethylformamide). The printing and curing sequence to produce near-net-shape composites is shown in Fig.~\ref{Fig01}a. To fabricate composite solids, first, the gel is extruded from the nozzle of a 3D printer at constant pressure $P$, while the print bed moves at a print speed $v$ relative to the extruder. The geometry of the nozzle and the print speed determine the shear rate and total shear strain imposed on the gel during this process. The shear deformation of the gel results in partial and yet irreversible degradation of the network of hydrogen bonds connecting the CNCs, thus allowing the gel to flow \cite{Rao2019}. However, by selecting a suitably high CNC concentration, the gel retains its shape after extrusion. The printed gel is then dried in the air with the print bed temperature set to 50$^{\circ}$C to extract the solvent, before being crosslinked. After drying, the samples are placed in an oven for a two-stage thermal cure, first at 80$^{\circ}$C and then at 130$^{\circ}$C, to fully cure the composite. Upon curing, the hydroxyl groups on the CNCs form covalent crosslinks with the epoxide monomers via a ring-opening mechanism \cite{Khelifa2016}. A photograph of the printing apparatus is shown in Figure~\ref{Fig01}b, and further details on gel formulation and printing are provided in supplementary Figure~\ref{composite_formulation_printing}.   

The shear history applied during the extrusion process plays a key role in regulating the grain size in the composite. Moreover, controlling the layer thickness is important to limit shrinkage stresses and the formation of macroscopic defects during solvent evaporation from the gel. Thus, additive manufacturing is a desirable process for the synthesis of bulk composites from the gel precursors, and potentially enables control of the micro and macro structure of the composite. Herein, we refer to the CNC gels as ``CG". We also discuss the processing of a UV-curable gel, denoted ``UV-CG", which is identical to the CGs, except for the addition of a photoinitiator, which promotes the same crosslinking mechanism between CNCs and epoxide upon UV exposure. Both the CG and UV-CG were printed using the direct-write process, yet the UV-CG is exposed to ultraviolet light after printing, followed by heating to complete crosslinking (see supplementary information section 1).

\begin{figure*}
    \centering
	\includegraphics[width=1.3\columnwidth]{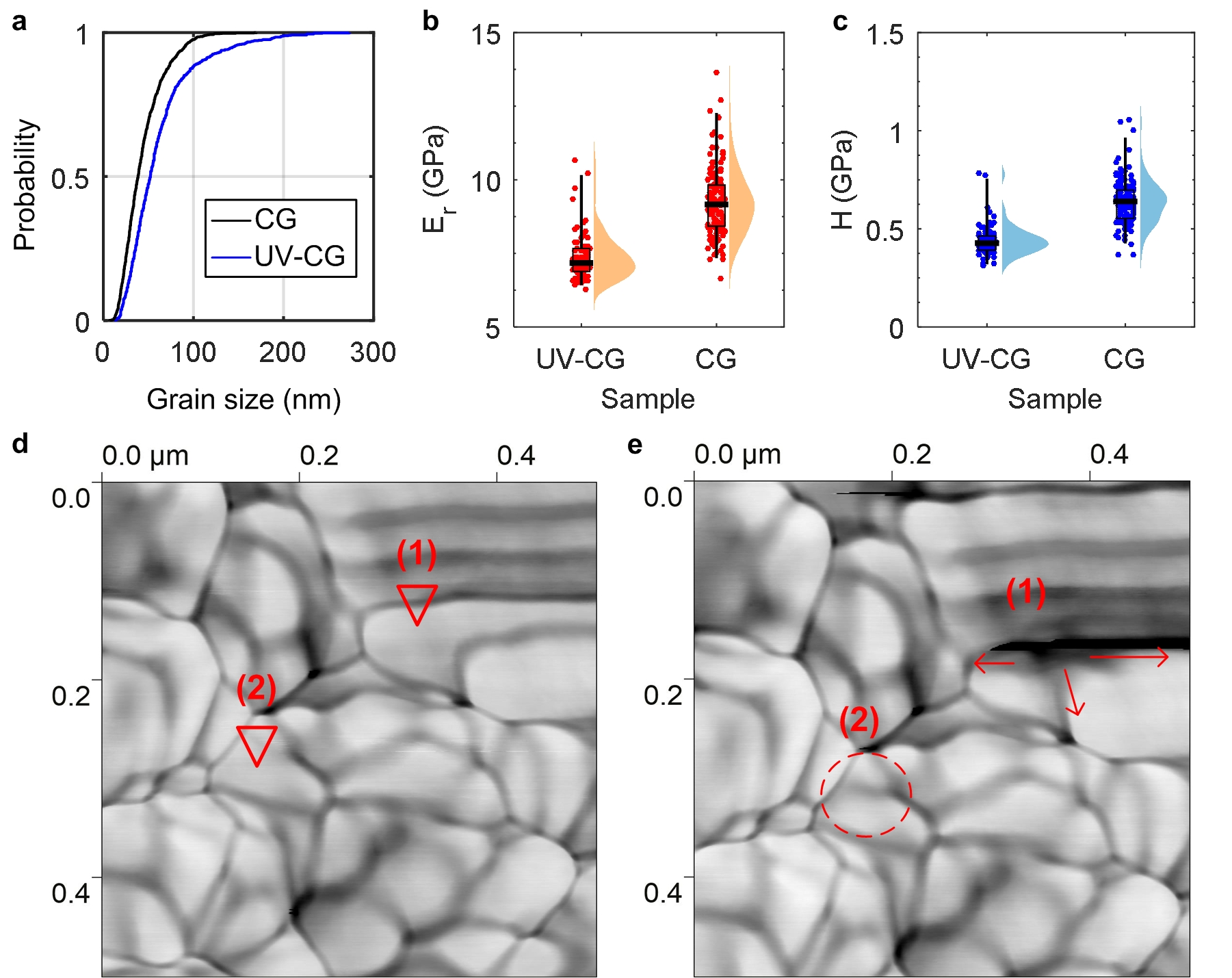}
	\caption{Influence of grain size on micromechanical properties of CNC-epoxy composites: a) Cumulative distribution of grain sizes in UV-cured composites (UV-CG) and thermally cured composites, without crosslinkers (CG). b) Indentation modulus $E_{\rm r}$ and c) Hardness $H$ determined by nanoindentation, with 63\% wt.~CNCs in both cases. Data are presented as a ``raincloud" plot \cite{Allen2018}; the box represents one standard deviation above and below the mean, and the whiskers represents the 75\% confidence interval. Each data point represents an indent at a distinct location on the sample. d) AFM images of a UV-cured composite (UV-CG) showing highly variable grain sizes. The triangles mark the locations where two indentation tests were performed, indicating large (1) and small (2) grains, respectively. e) Image of the same region after indentation, showing that the indent in location (1) has triggered a crack larger than 150~nm at the interface between two adjacent grains, whereas the indent in location (2) had a localized impact (crack length $<50$~nm).}
	\label{Fig02}
\end{figure*}

The process of dispersion of CNCs in the preparation of the gel is critical to avoid precipitation of large aggregates of CNCs and phase separation. In the present work, a probe sonicator is used to disperse CNCs in the solvent. The sonication process was optimized for sample volumes of around 10~mL. Furthermore, printing samples beyond the centimeter scale requires apparatus such as an enclosed print chamber to control solvent evaporation and a large-volume pneumatic extruder to print a larger volume of gels. While these limitations preclude large-scale mechanical testing, we comprehensively study the structure-property relationships from nanometer to millimeter scale, capturing the interactions between clusters of a few CNCs and relating it to the ``composite'' behavior of the bulk material. The methods of micro-hardness and fracture toughness testing applied here are routinely used to characterize materials with volume-dependent heterogeneity \cite{Borodich2015}. These methods can serve as useful approximations of conventional engineering tests, while providing valuable micro structural insights for materials with multiple length scales \cite{Akono2011,Akono2012,Randall2009}. 

 A preliminary examination of the CNC-epoxide composite, by comparing the optical and cross-polarized images, shows the presence of nematic domains of CNCs (Figures \ref{Fig01}c and \ref{Fig01}d). The bright colors in the cross-polarized images have been widely used to identify nematic liquid crystals formed in CNC suspensions \cite{Abitbol2014,Schutz2015,Gray2015}. AFM amplitude and phase images within these domains, reveal an anisotropic, nanoscale grain structure of typical size 100~nm, as shown in Figure~\ref{Fig01}e and \ref{Fig01}f respectively. As observed in the amplitude image, the grain orientation indicates a local order corresponding to the direction of extrusion. Furthermore, upon comparing the amplitude and phase images, we find that the phase recorded for the grains is relatively constant, regardless of the topographical variations across the scanned region. However, the grain boundaries remain distinct in phase throughout the image. This observation suggests that the composition and mechanical properties of the grain boundaries are distinct from those of the grains.
 
 \begin{figure*}
    \centering
	\includegraphics[width=1.3\columnwidth]{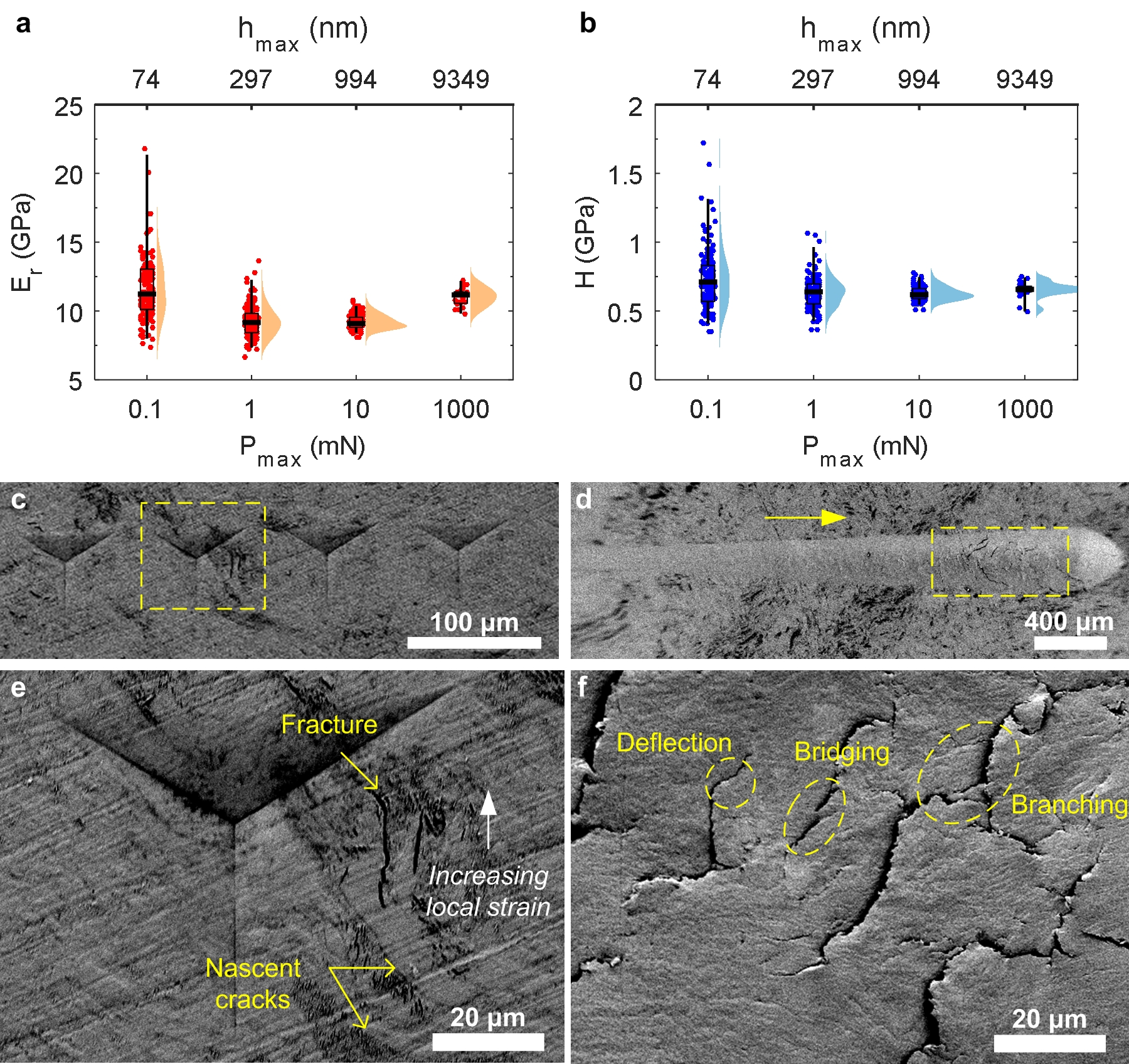}
	\caption[Multiscale mechanical properties of thermally-cured CNC composites]{Multiscale mechanical properties of thermally-cured CNC composites with 63\%~wt.~CNCs: a) Elastic modulus and b) Hardness measured by nanoindentation at different peak loads ranging between 0.1~mN and 1~N. The corresponding average indentation depths range from 74~nm to 10~$\mu$m. For each set of data, the box plot indicates the mean, 75\% confidence intervals, and maximum range of the data. c) SEM image showing the plastic deformation due to an indent at a peak load of 1000~mN, d) Plastic deformation resulting from a scratch test performed under increasing normal load from 30~mN to 3~N. The arrow indicates the scratch direction. High magnification images of the e) indent and f) scratch showing brittle plasticity at the microscale, and the fracture toughening mechanism contributing to bulk ductility.}
	\label{Fig03}
\end{figure*}

Such a distinction is due to the microstructure of the gel and how it evolves during processing. The CNC-polymer gels possess a glassy microstructure composed of densely packed clusters made of CNCs linked by hydrogen bonds \cite{Rao2019}. These clusters are partially broken down during the printing step, as the shear flow irreversibly breaks hydrogen bonds. Once the gel is brought to rest, the clusters can re-aggregate by van der Waals attraction that is counteracted by electrostatic repulsion between the CNCs, due to their native negative surface charge \cite{Moon2011}. As a result, the composite contains residual clusters forming larger grains, as well as freely dispersed individual CNCs surrounded by the epoxide monomer, as shown in Fig.~\ref{Fig01}d. The extent to which the shear flow disrupts the initial gel microstructure depends on the total shear strain applied during extrusion.

The local Young modulus of the composite is directly related to the strength of the intermolecular interactions at the CNC-CNC and CNC-polymer interfaces. Bimodal AFM imaging shows a modulus of 50~GPa at the grain boundaries and 30~GPa within the grains (see supplementary Fig.~\ref{grain_boundaries_modulus}). The grains are composed of tightly packed and hydrogen-bonded CNCs and are likely to be infiltrated to a lower degree by the monomer, particularly as they persist during extrusion. As a result, the grains may have a lower degree of covalent crosslinking than the grain boundaries and thus a lower modulus.

 The cumulative grain size distributions of the thermally cured (CG) and UV cured (UV-CG) composites were determined using AFM images of three regions on each sample (see supplementary Figs.~\ref{hv_grain_boundaries}, \ref{H070_grain_sizes} and Table~\ref{H070_lognorm_parameters}). We find a significant difference in grain size distributions for composites prepared from gels with (UV-CG) and without (CG) the photoinitiator, despite having the same CNC concentration. In the UV-cured gels, the photoinitiator has the effect of screening the surface charges on the CNCs, promoting aggregation and grain growth \cite{Rao2019}.  As a result, UV-cured composites have a larger average grain size with broader distribution, including grains larger than 200~nm. In contrast, thermally-cured composites have a narrow grain size distribution with an average size around 100~nm (Fig.~\ref{Fig02}a). The indentation modulus (or reduced modulus) $E_{\rm r}$ and hardness $H$ of the composites were obtained by nanoindentation, and data is shown in Fig.~\ref{Fig02}b and \ref{Fig02}c, respectively. Nanoindentation grids, consisting of 12$\times$12=144 indents, were performed on each sample to a typical depth of 1~$\mu$m, corresponding to a peak load of 10~mN (see supplementary Fig.~\ref{auto_outlier_stats} and corresponding text for further discussion on the nanoindentation data analysis). The thermally-cured composite has distinctly better mechanical properties than the UV-cured composite, ($E_{\rm r}=9.2$~GPa vs.~7.3~GPa, and $H=0.64$~GPa vs.~0.44~GPa, respectively). 

The broader grain size distribution of UV-cured composites suggests that grain growth during the printing and curing processes must be controlled to achieve improved mechanical properties. To illustrate this point, we compare indentations performed with AFM tips in two representative locations of a UV-cured composite (Fig.~\ref{Fig02}d). Location~(1) is chosen in a region with grains larger than 200~nm, whereas location (2) is picked in a region with smaller grains. The indents performed in these two regions are visible in Fig.~\ref{Fig02}e. Both indents were force-controlled, with a peak load of 0.5~$\mu$N. In region (1), the cracks emanating from the vertices of the indent are observed to propagate along the grain boundaries, resulting in the brittle fracture of the composite over a length scale greater than 150~nm. By contrast, the plastic deformation in region (2) is confined to a region smaller than 100~nm in diameter, and the extent of the corresponding crack is restricted by the grain boundaries. Therefore, a narrow grain size distribution with a small average grain size is ideal for improving resistance to long-range crack propagation. 
We systematically varied the CNC mass fraction, finding that a CNC content of about 63\% wt.~corresponds to the maximum values of indentation modulus and hardness (see supplementary Fig.~\ref{printed_cnc_mass_fraction}). The emergence of an optimum concentration may be attributed to the fact that a sufficient amount of polymer is necessary to surround and crosslink the CNCs to one another. Indeed, at higher CNC concentrations, the crosslink density is low, and the composite is more brittle, hence less resistant to plastic deformation. By contrast, at lower CNC concentrations, the CNC contribute less to the composite modulus.

\begin{figure}
    \centering
	\includegraphics[width=0.8\columnwidth]{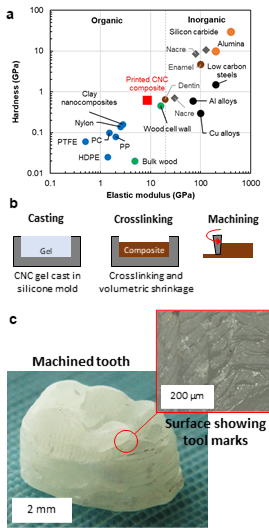}
	\caption[Ashby plots]{Mechanical properties of the printed CNC composites in comparison with those of natural and synthetic materials: \cite{ashby2011,Gao2017,Shen2004,Paplham1995,Tze2007,Briscoe1998,Song2018,Lichtenegger2002,Imbeni2005,Staines1981,Tesch2001GradedDentin,Song2015HierarchicalShell,Barthelat2006MechanicalPerformance,Bruet2005NanoscaleNiloticus} a) Average values of elastic modulus versus hardness determined by nanoindentation  and b) Fabrication sequence for a micromachined composite parts c) Model of a tooth machined from a block of CNC-epoxide composite, with inset showing machined surface.}
	\label{Fig04}
\end{figure}

At this point, we can draw an analogy between the microstructure and mechanics of CNC-epoxy composites and those of biological composites such as bone and nacre, which are largely composed of brittle constituents, and yet display ductile mechanical behavior at larger scales \cite{Barthelat2016}. Here, statistical results from nanoindentation can connect the nanoscale mechanical behavior to the micro-mechanical properties of the CNC-epoxy composites, by relating the indentation depth to the intrinsic length scale of the microstructure of the material \cite{Kushch2015}. Indentation at shallow depth allows individual phases in the sample microstructure to be resolved, while deep indents provide averaged values representative of the macroscopic mechanical properties. We performed grids of 144 indents on a composite with 63\%~wt.~CNCs, at peak loads, $P_{{\rm max}}$ from 0.1~mN to 1000~mN, which correspond to average indentation depths, $h_{\rm{max}}$ ranging from ($74\pm$11)~nm to ($9349\pm$355)~nm respectively. The values of $E_{\rm r}$ and $H$ are reported as a function of both $P_{\rm max}$ and $h_{\rm max}$ in Figure~\ref{Fig03}a and \ref{Fig03}b respectively. Both the indentation modulus and the hardness of the CNC composites are observed to be independent of depth from the nanoscale to the microscale, with overall averaged values of $\bar{E_{\rm r}}=$10.3~GPa and $\bar{H}=0.66$~GPa respectively. The Young modulus can be calculated using the indentation modulus and Poisson's ratio \cite{Oliver1992}, here giving $E$ = 8.5~GPa. For the chosen (Berkovich) indenter geometry, the length scale of the sample probed by an indent of depth $h_{{\rm max}}$ is approximately $3h_{{\rm max}}$ \cite{Constantinides2006}.

In general, the measured hardness increases with decreasing indentation depths due to the effect of discrete dislocations and surface roughness, and converges to the intrinsic hardness of the material at sufficiently large indentation depths. For heterogeneous materials, large indentation depths may be necessary to probe the ``composite'' properties, rather than those of individual phases \cite{Pharr2010,ShellDeGuzman1993,Miller2008}. For the CNC-epoxy composites, we observe that the hardness and elastic modulus measured over indentation depths spanning from around 100~nm to 1~$\mu$m are comparable (Fig.~\ref{Fig03}b). Therefore, we can conclude that a volume with a critical dimension of around 100~nm, comprising a few grains and their associated interfaces, is a representative element of the composite. The constant hardness of the CNC-epoxy composites versus indentation depth illustrates that toughening mechanisms enabled by the CNCs are hierarchical and contribute improving the mechanical properties of the material at length scales that are several orders of magnitude larger than the CNC particles.

We now turn to the analysis of the plastic deformation after microindentation, along with scratch testing, to further quantify the fracture resistance of CNC-epoxy composites. A conical indenter is moved at constant speed across the surface of the sample while steadily increasing the vertical load and recording the tangential force $F_T$ exerted on the indenter. Once normalized by the contact area, $F_T$ converges, for large enough penetration depth of the indenter, towards the fracture toughness; prior studies show this matches the plane strain fracture toughness from conventional 3-point bending tests on notched samples \cite{Akono2011,Akono2014}. Both microindentation and scratch tests generate fracture patterns in the sample over micron to millimeter length scales.
In Figure~\ref{Fig03}c we show the residual plastic deformation from microindentation at a peak load of 10~mN. We note the absence of linear cracks emanating from the indent's vertices, indicating that the composite plastic behavior is ductile, rather than brittle \cite{Skrzypczak2009}. Moreover, the absence of linear cracks is also consistent with the amorphous microstructure of the composites \cite{Cook1990}. Figure~\ref{Fig03}d shows the plastic damage observed after a scratch test, wherein the arrow indicates the direction of the scratch test. The vertical load during the scratch test is increased linearly from 30~mN to 30~N along the scratch direction. For loads larger than 28~N, we observe cracks perpendicular to the scratch direction. This pattern of damage together and the absence of parabolic crack patterns at low vertical loads are characteristic of ductile polymers \cite{Jiang2009}. From the scratch load versus depth (Supplementary Fig.~\ref{CSA214_scratch_Kc}), we estimate a fracture toughness of 5.2~MPa.m$^{1/2}$ for a composite with 63\% wt.~CNCs.

High magnification images for the indent and the scratch shown in Figures~\ref{Fig03}e and \ref{Fig03}f, respectively, indicate that cracks originate from the regions subjected to large strains around the indenter. These cracks ultimately grow and branch into larger cracks, causing microscale fractures. The fracture path is not continuous, indicating that fracture energy is distributed by crack deflection and the splitting of cracks into multiple separate segments. Examination of the sample after the scratch test suggests that the crack propagation mechanisms are observed over longer length scales compared with those in the indent (Figure~\ref{Fig03}f). In addition to being deflected, the propagation of long individual cracks is interrupted by branching and bridging. In the latter case, a bridge of material splits the crack into multiple regions. The crack geometry and propagation are analogous to that observed for nacre \cite{Bouville2014,Gao2017,Niebel2016133}, and are a direct consequence of the deflection and arrest of cracks by the glassy, nanoscale grain structure of the CNC-epoxy composites. The fracture energy is dissipated at the grain boundaries due to the high crosslink density and the discontinuous arrangement of the grains.

Our results show that the micromechanical properties of the CNC-epoxy composites exceed those of many engineering polymers and all-organic materials, and that they are comparable to those of wood cell wall. (Figure~\ref{Fig04}a) \cite{ashby2011,Gao2017,Shen2004,Paplham1995,Tze2007,Briscoe1998,Song2018,Lichtenegger2002,Imbeni2005,Staines1981}. The reported modulus and hardness of nacre vary significantly based on source \cite{Barthelat2006MechanicalPerformance,Song2015HierarchicalShell,Bruet2005NanoscaleNiloticus}. While the predominantly inorganic phase of nacre results in a higher stiffness, the hardness of the CNC-epoxy composites is comparable to certain species of nacre. Notably, we observe these mechanical properties over length scales from 10 to $10^3$ times the diameter of the CNCs. Wood, which is the source of CNCs, has a hierarchical structure wherein the microscale cell walls comprise layers of aligned cellulose microfibrils, interconnected by cellulose nanofibrils \cite{Moon2011,Kretschmann2003}. Due to its macroscale cellular structure, the modulus and hardness of wood measured by microindentation are significantly lower than those measured by nanoindentation \cite{Moon2009}. Indeed, the hardness of a wood cell wall measured by nanoindentation is around 0.4~GPa, whereas the bulk hardness of wood, measured at depths up to 500~$\mu$m, ranges between 60 and 90~MPa.  

Printed CNC-epoxy composites can also be further processed to produce macroscale objects having complex shapes. Here we demonstrate the fabrication of a model of a tooth, a potentially compelling application for a hard and tough biocomposite, by molding and curing a block of the CNC gel (Figure~\ref{Fig04}b), followed by micromilling (details in supplementary information). The machined model of a tooth, shown in Figure~\ref{Fig04}c illustrates that the bulk stiffness and the ductility of the CNC-epoxy composite enable machining of complex surface contours. 

\section{Conclusion}
We have demonstrated the processing and multi-scale mechanics of CNC-epoxy composites with high CNC mass fractions, derived from versatile gel-based precursors. These composites comprise grains of aggregated CNCs that are crosslinked by covalent bonding. The tortuous paths formed by the grain boundaries resist the linear propagation of cracks between grains. The bridging, deflection, and branching of cracks impart macroscale toughness to the CNC-epoxy composites. The stiffness, hardness, and fracture toughness of the CNC composites exceeds those of many engineering polymers. The composite response is ductile despite the inherently brittle behavior of the CNC grains. Our findings suggest that the addition of significant fractions of CNCs to petroleum-based polymers can potentially improve the mechanical properties and reduce the environmental impact of these ubiquitous materials. The fracture toughening mechanisms observed in these composites, as well as the relationships established between synthesis and microstructure are more broadly applicable to molding and additive manufacturing of other nanocrystalline-polymer composites.\\

\acknowledgements
The authors thank Sooraj Narayan and Prof.~Lallit Anand for assistance with the finite element model. Financial support was provided by the Procter and Gamble Corporation, and we thank Neville Sonnenberg for discussions related to the project. C.E.O. was supported by the United States Department of Defense (DoD) through the National Defense Science \& Engineering Graduate Fellowship (NDSEG) Program.

\bibliography{references}

\renewcommand{\thefigure}{S\arabic{figure}}
\setcounter{figure}{0} 

\clearpage
\section{Supplementary Information}

\section{Direct ink writing}
\begin{figure*}
	\includegraphics[width=1.3\columnwidth]{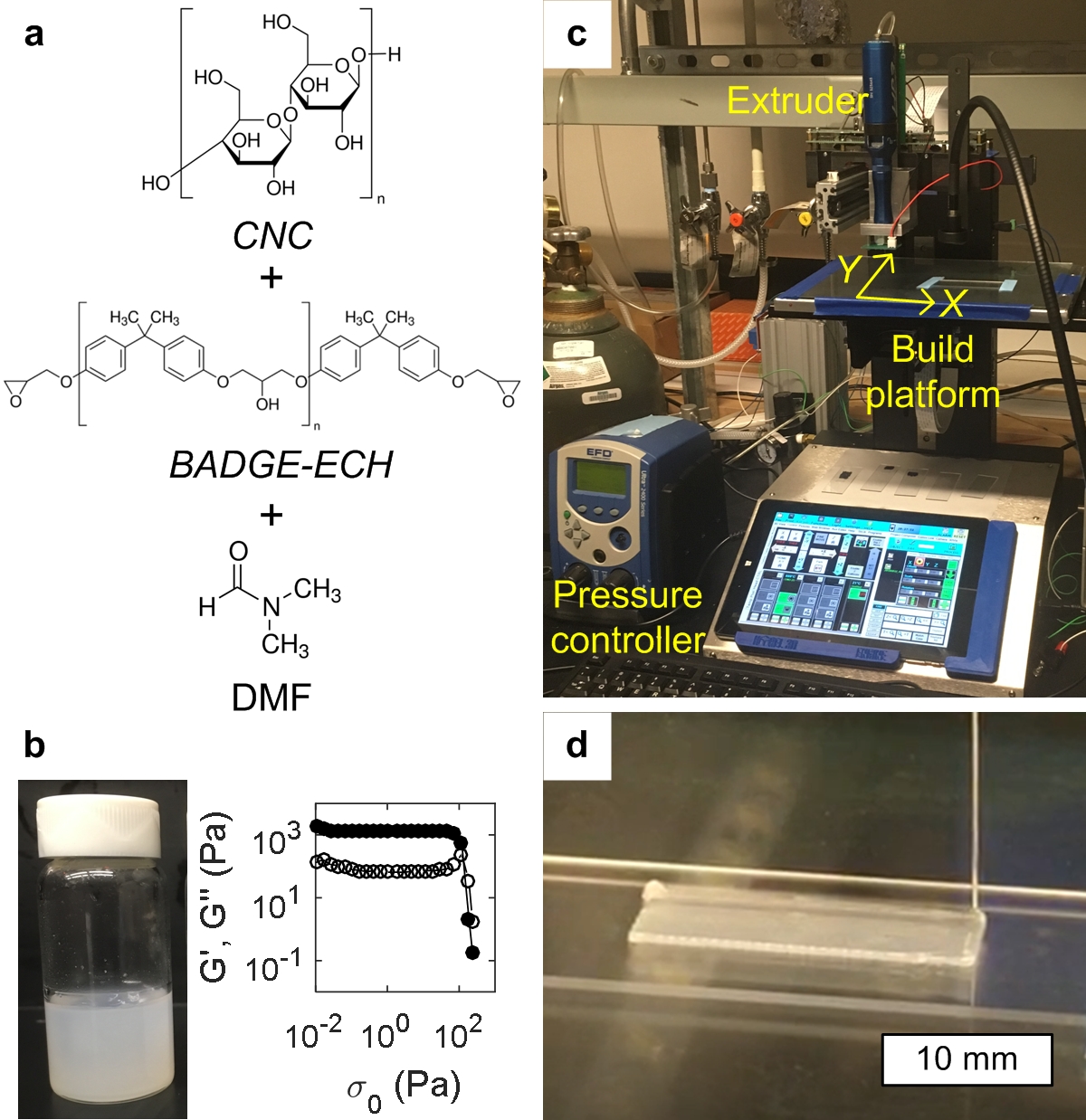}
	\caption[Direct-write printing of CNC composite gels]{Formulation and direct-write printing of crosslinked CNC composites: a) Chemical formulation of the gel, b) Photograph showing a sample of composite gel with 12.5\% wt.~CNCs, and a rheological test (stress-sweep at $f=1$~Hz) illustrating the existence of a yield stress, c) Hyrel 3D printer with a pneumatic extruder and d) Photograph showing an extruded layer during printing onto a glass slide.}
	\label{composite_formulation_printing}
\end{figure*}

The gels are composed of cellulose nanocrystals (CNCs) and an epoxide oligomer dispersed in an organic solvent, here dimethylformamide (Fig.~\ref{composite_formulation_printing}a). \textit{UV curable CNC gels} (UV-CG) were prepared by first dissolving an epoxide oligomer, a photoinitiator, and a thermal crosslinker in dimethylformamide (DMF). The oligomer is poly(bisphenol-a-co-epichlorohydrin) diglycidyl ether ($M_n\sim355$ g/mol), the cationic photoinitiator is triarylsulfonium hexafluorophosphate (50\% in propylene carbonate) and the thermal crosslinker is 4-aminophenyl sulfone. All reagents were obtained from Sigma Aldrich. Freeze-dried CNC powder (Cellulose Lab, New Brunswick, Canada) is added to the mixture and then dispersed by probe sonication at 40\% amplitude for 15 seconds (750 W Sonics Vibra Cell). Reference \textit{composite gels} (CG) were prepared using the same process, with the exception of adding the photoinitiator and thermal crosslinker. In all the gels, the mass ratio of the epoxide oligomer to the CNCs is 10\%, and the mass of the cationic photoinitiator is 10\% of the mass of the epoxide oligomer. The amine to epoxide molar ratio is 0.4. The mass fractions of the crosslinkers are selected to allow UV curing of millimeter-thick layers within a few minutes, and accelerated thermal curing at lower temperatures. This process closely follows the gel formulation described in a previous study.\cite{Rao2019}

The hydroxyl groups on the CNCs crosslink with the epoxide monomers upon heating, which leads to the formation of a gel.\cite{Khelifa2016} A stress sweep performed at frequency $f=1$~Hz on a gel of composition with 12.5\% CNCs by mass relative to solvent, placed in a parallel-plate geometry connected to a stress-controlled rheometer (AR-G2, TA Instrument) is reported in Fig.~\ref{composite_formulation_printing}b. Note that in the gel, epoxide monomer is dispersed in the solvent in a ratio such that the final mass fraction of CNCs in the composite, is 63\% after the solvent is dried . The linear mechanical properties of the gel are dominantly elastic ($G'_0=1000$~Pa $\gg G''=$100~Pa), and the gel yields beyond a certain critical stress of about $\sigma_{\rm c}\simeq 100$~Pa. 

A commercially available desktop 3D printer Hyrel Engine SR printer (Hyrel 3D, Norcross GA) was modified for direct ink writing of crosslinked CNC composites by replacing the standard syringe extruder with a Nordson HP3 pneumatic syringe extruder (Fig.~\ref{composite_formulation_printing}c). 
The extruded composite gel is pictured in Fig.~\ref{composite_formulation_printing}d. The printed bed is heated to 50$^{\circ}$C, and the gel is deposited in multiple layers through a nozzle of diameter 0.3~mm. The gel expands slightly upon extrusion, and the thickness of each layer is approximately 0.5 mm. UV-curable gels are partially cured under a UV flood lamp (Dymax 2000-EC) for 5 minutes. Following this, they still contain solvent, but have higher mechanical integrity. Aside from this step, the curing process is identical for both UV and thermally curable samples. A two-stage thermal cure - first at 80$^{\circ}$C for 6 hours and 130$^{\circ}$C for 4 hours is used to fully harden the samples. The printed samples are then polished on a rotary polisher using a silicon carbide polishing paper of grit P4000 (Buehler) prior to any mechanical testing. The thickness of the cured samples was around 1~mm, around 100 times the largest indentation depth.

\begin{figure*}
	\includegraphics[width=1.3\columnwidth]{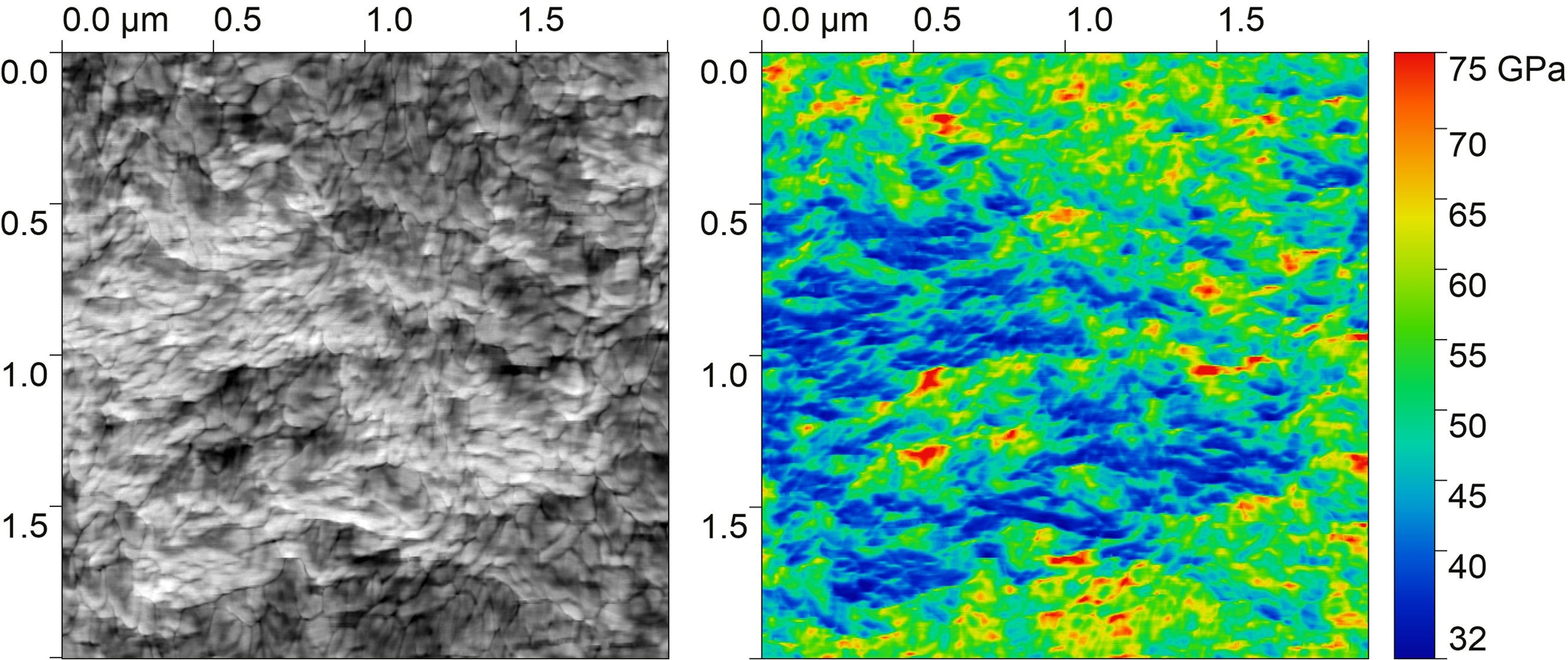}
	\caption[AMFM imaging of CNC-epoxide composites]{Visualization of grains and modulus map obtained by AMFM imaging}
	\label{grain_boundaries_modulus}
\end{figure*}

\section{Bimodal atomic force microscopy}
In traditional atomic force microscopy (AFM) a cantilever is excited near its resonant frequency, and the amplitude and phase of the cantilever are measured as the tip scans across the sample surface. This information is used to map the topography of the sample using Hertizan mechanics.\cite{Labuda2016} In bimodal AFM, the tip is excited at two separate eigenfrequencies. Amplitude feedback on the first eigenmode is adjusted to maintain a pre-determined set point. This method is used to determine topography information similar to conventional AFM. A frequency feedback loop operates on the second eigenmode to keep it at resonance as the sample is scanned. The measured changes in frequency are then used to calculate an interaction storage modulus and an effective indentation depth.\cite{Kocun2017} This method is known as \textit{AMFM imaging}, and provides simultaneous structural and mechanical information at lengths scales smaller than 100~nm, which are inaccessible by other techniques such as nanoindentation and scanning electron microscopy. Fig.~\ref{grain_boundaries_modulus} shows an AMFM image of the printed CNC-epoxide composites. Here the modulus map is obtained by first scanning the composite sample, obtaining the modulus with arbitrary units. Then the same tip and cantilever gain and height settings are used to scan a bare silicon wafer. A calibration factor is then calculated assuming a modulus of 130~GPa for silicon,\cite{Hopcroft2010} and is in turn used to determine the modulus values of the CNC composite.

\begin{table*}[h!]
    \centering
    \begin{tabular}{ | m{2.5cm} | m{2.5cm}| m{2cm} | m{2.5cm} | m{1.5cm} |} 
    \hline 
    Region & $\mu$ & $\sigma$ & $\bar{x}$ (nm) & $M$ (nm) \\ 
    \hline \hline 
    a & 4.08 & 0.5 & 66.81 & 58.97  \\ 
    \hline
    b & 3.83 & 0.52 & 52.69 & 46.03  \\ 
    \hline
    c & 3.94 & 0.54 & 59.48 & 51.46  \\ 
    \hline
    \end{tabular}
    \caption{Lognormal distribution fit parameters for the grain sizes distribution displayed in Figure \ref{H070_grain_sizes}. Data include the scale parameters, $\mu$ and $\sigma$ as well as the mean, $\bar{x}$ and median, $M$.
    }
    \label{H070_lognorm_parameters}
\end{table*}

\begin{figure*}
	\includegraphics[width=1.3\columnwidth]{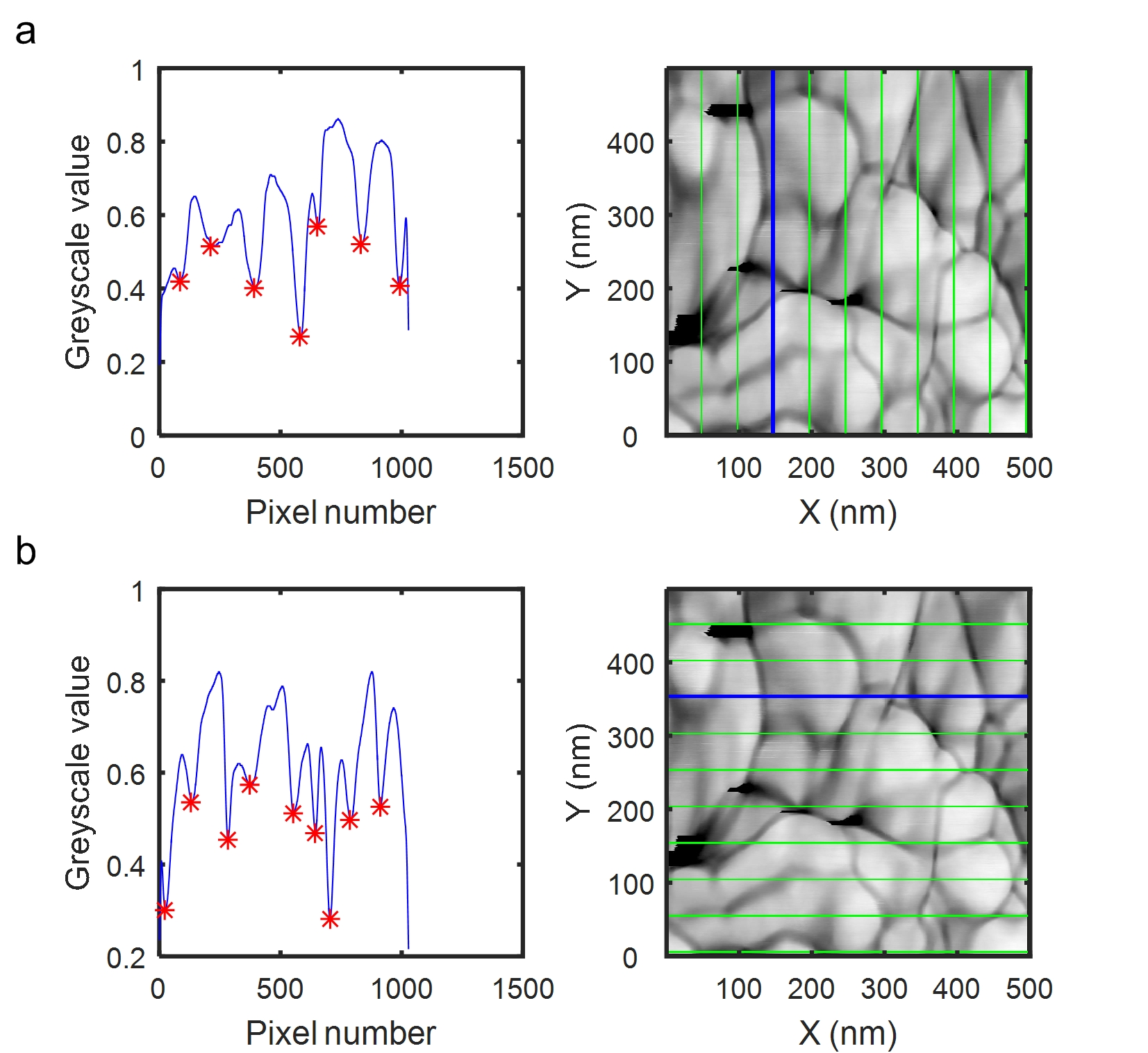}
	\caption[Horizontal and vertical grain boundary detection]{Grain boundary detection along a) horizontal and b) vertical directions. For each data set, the right image correspond to the AMFM data, while the left greyscale profile corresponds to the blue line in the image.}
	\label{hv_grain_boundaries}
\end{figure*}

\begin{figure*}
	\includegraphics[width=1.3\columnwidth]{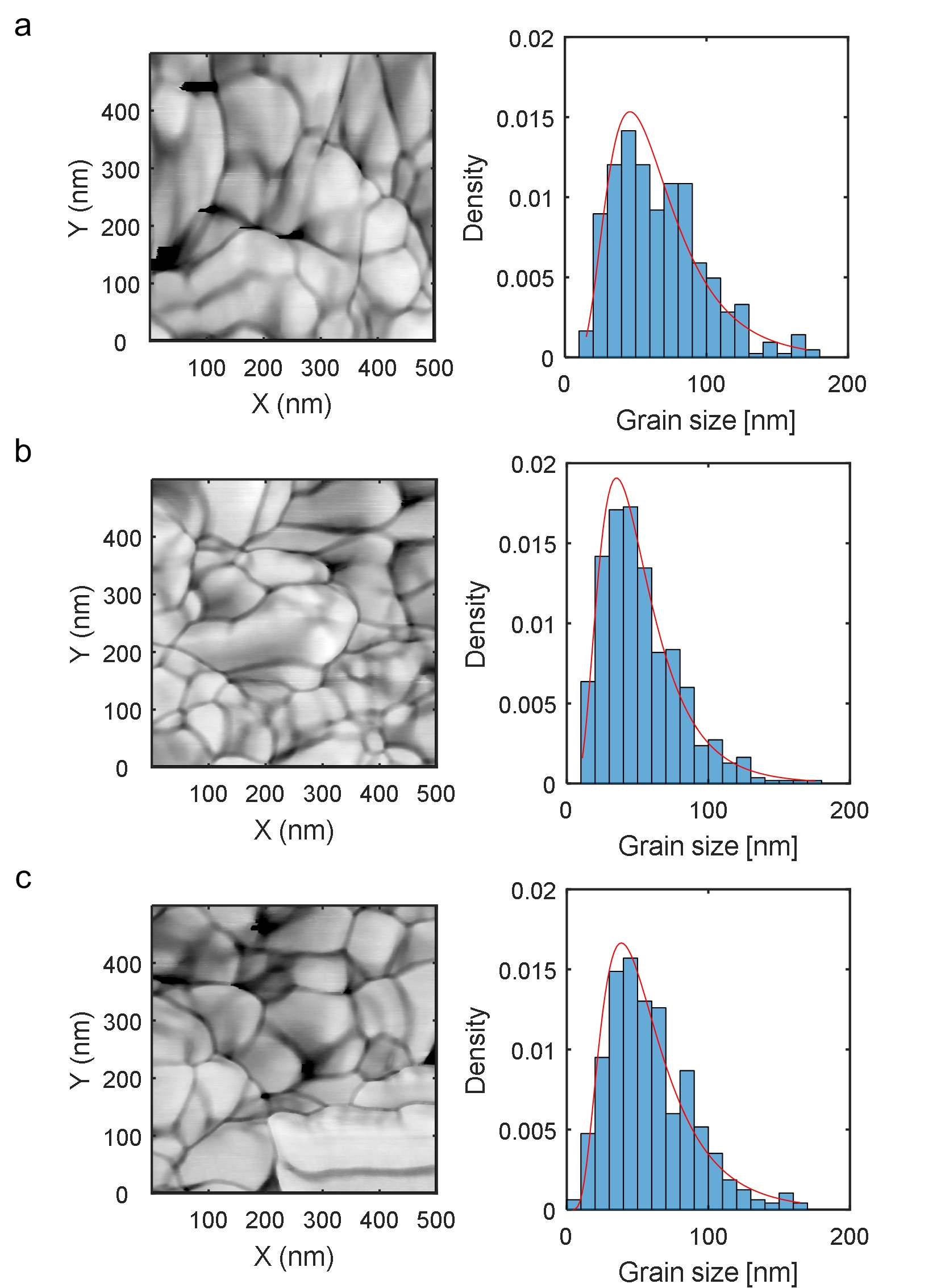}
	\caption[AMFM image and Grain size probability distribution at three different location of a composite sample]{AMFM image (left) and grain size probability distribution (right) at three different locations labelled a, b and c of a CNC-polymer composite sample, with 63\% wt. CNCs. The red curves on the probability distribution graphs correspond to the best fit of the data by a lognormal function of parameters listed in Table~\ref{H070_lognorm_parameters}}
	\label{H070_grain_sizes}
\end{figure*}

\section{Grain size measurement}

Automated measurement of grain sizes from binary images is performed using the linear intercept method, according to the ASTM E1382 standard.\cite{Lehto2014, ASTME1382-972004} The measurement process involves plotting evenly spaced lines across the image, and recording the distances between the detected grain boundaries along each line. This measurement is repeated in four directions, 0$^{\circ}$, 90$^{\circ}$, 45$^{\circ}$, 135$^{\circ}$. The measurements along each direction are stored in separate variables and finally combined into a single histogram. 

Grain size distributions are computed from the AMFM images. Conventionally, a threshold is used to convert the greyscale image into a binary (black and white) image. The grains and grain boundaries have opposite binary values, and can easily be detected. A Gaussian blur is first applied to the raw indentation image, serving as a low-pass filter to reduce noise. However, the grain boundaries in the indentation images, while  visible to the eye, have varying greyscale values. A fixed threshold, therefore, would not accurately detect the grain boundaries across the entire image. Rather than using a fixed threshold, the grain boundaries are identified as the local minima of the greyscale values along a given measurement trace. Fig.~\ref{hv_grain_boundaries} shows lines along which grain sizes are measured in the 0$^{\circ}$ and 90$^{\circ}$ directions. The normalized greyscale values along the blue highlighted trace are plotted in Fig.~\ref{hv_grain_boundaries}a, showing the local minima (red stars) corresponding to the grain boundaries. This process is repeated for diagonal traces along the 45$^{\circ}$ and 135$^{\circ}$ directions. 
The local minimum method is adaptive to account for variations in the greyscale value of the grain boundaries within a single image. Furthermore, this method does not require any inputs that could cause inconsistencies across multiple images. 

Following the ASTM E1382 protocol, the grain boundaries measured along each direction are stored in separate variables and finally combined into a single histogram. Figure~\ref{H070_grain_sizes} shows a probability distribution of the combined dataset for three regions, denoted (a), (b) and (c) of the same sample alongside the corresponding indentation maps. The three regions are distant by about 1~mm from each other. The probability distributions of the grain sizes are well described by a lognormal probability distribution. The fit parameters corresponding to the three regions are given in Table~\ref{H070_lognorm_parameters}. While the mean and median grain sizes are similar in all three regions, we observe that there are a few grains larger than 200~nm in each region. 

\begin{figure*}
	\includegraphics[width=1.3\columnwidth]{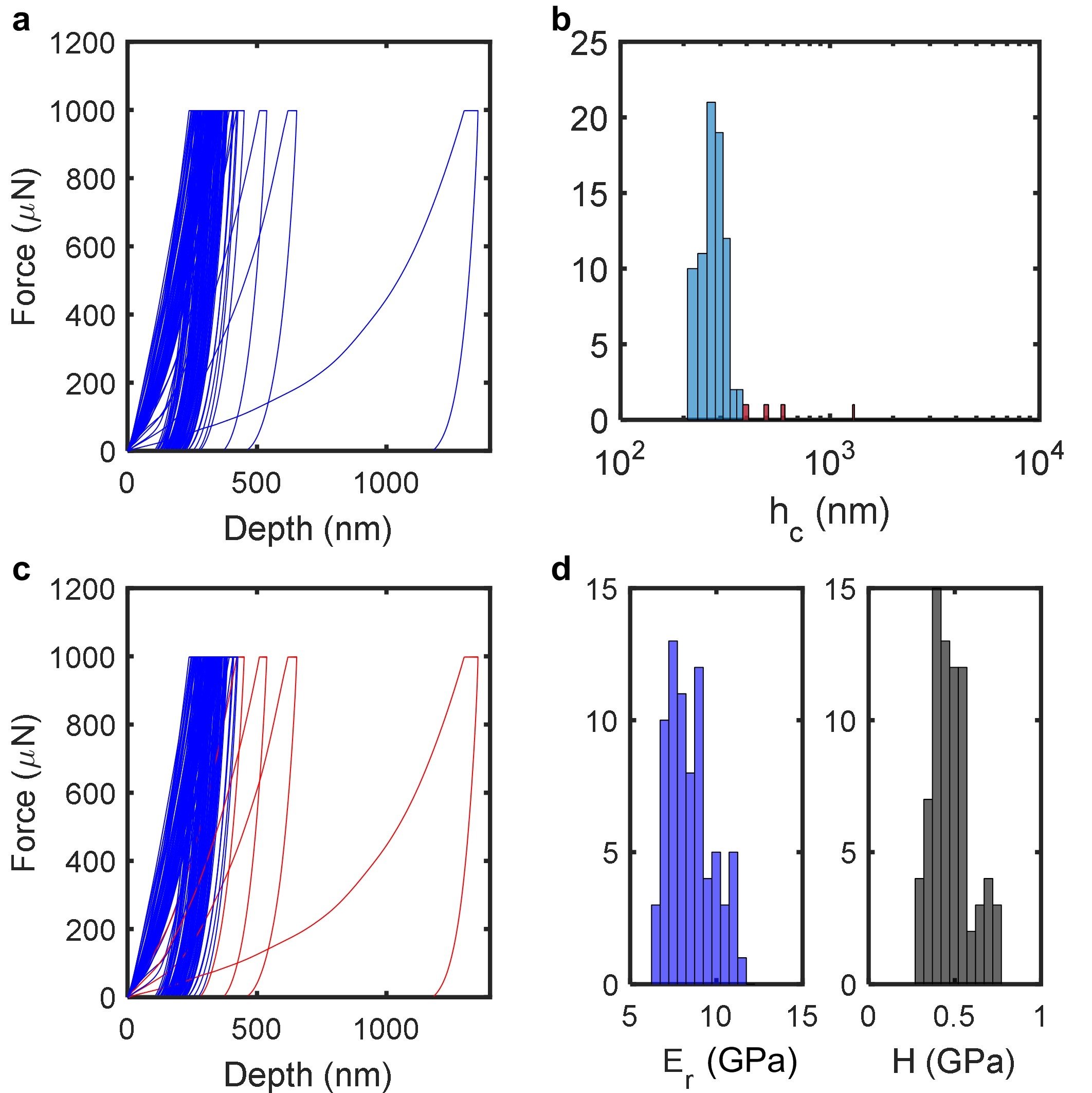}
	\caption[Data analysis from nanoindentation tests]{ a) Raw data load-displacement (P-H) curves obtained for 81 indents distant by 10~$\mu$m performed at a maximum force 1~mN on a composite with 78\%~wt.~CNCs, b) Histogram of the contact depth $h_{\rm c}$ showing outliers in red, c) Same P-H curves as in a) where the outliers are highlighted in red, d) Histograms of the indentation modulus $E_{\rm r}$ and the hardness $H$ without the outliers.}
	\label{auto_outlier_stats}
\end{figure*}

\section{Detection of outliers in the nanoindentation data}
Statistical nanoindentation tests were conducted on a Hysitron Triboindenter equipped with a three-sided pyramid diamond Berkovich indenter. A 3-step trapezoidal load profile is imposed with loading and unloading steps 10~s each, separated by a 5~s hold at peak load. The vertical position of the indenter is  recorded simultaneously. Each indent results in a load-displacement (P-H) curve, which is analyzed following the method of Oliver and Pharr to deduce the indentation modulus $E_{\rm r}$ and the hardness $H$ at the locus of the indent, over a volume with corresponds to 3 to 5 times the maximum indentation depth.\cite{Oliver1992}
Both systematic and non-systematic errors affect the load-displacement (P-H) curves obtained from a typical nanoindentation experiment. On the one hand, systematic errors are most often attributed to the detection of the contact point between the indenter and the sample surface.\cite{Xia2014} On the other hand, non-systematic errors are due to measurement noise, temperature fluctuations, external disturbances, and material variations.\cite{Menk2012} 

To detect and remove the outliers due to systematic errors, we have devised a simple algorithm. Figure~\ref{auto_outlier_stats}a, shows a set of P-H curves obtained from 81 force-controlled indents on a CNC-polymer composite with 72\%~wt.~CNCs. The indents were spaced 10~$\mu$m apart, to ensure sufficient separation of the plastic zones, such that each indent can be treated as a separate statistical event. The resulting distribution of contact depth is shown in Figure \ref{auto_outlier_stats}b. The histogram is divided into 25~nm bins, and considering that our samples are relatively homogeneous at the nanoscale, bins with less than two points are considered to be outliers. Using this method, three P-H curves corresponding to the outliers are detected (see red curves in Figure \ref{auto_outlier_stats}c). Finally, the remaining data points are used to calculate the histogram of the indentation modulus and the hardness (Fig.~\ref{auto_outlier_stats}d). 

\section{Additional nanoindentation data}
\begin{figure*}
	\includegraphics[width=1.2\columnwidth]{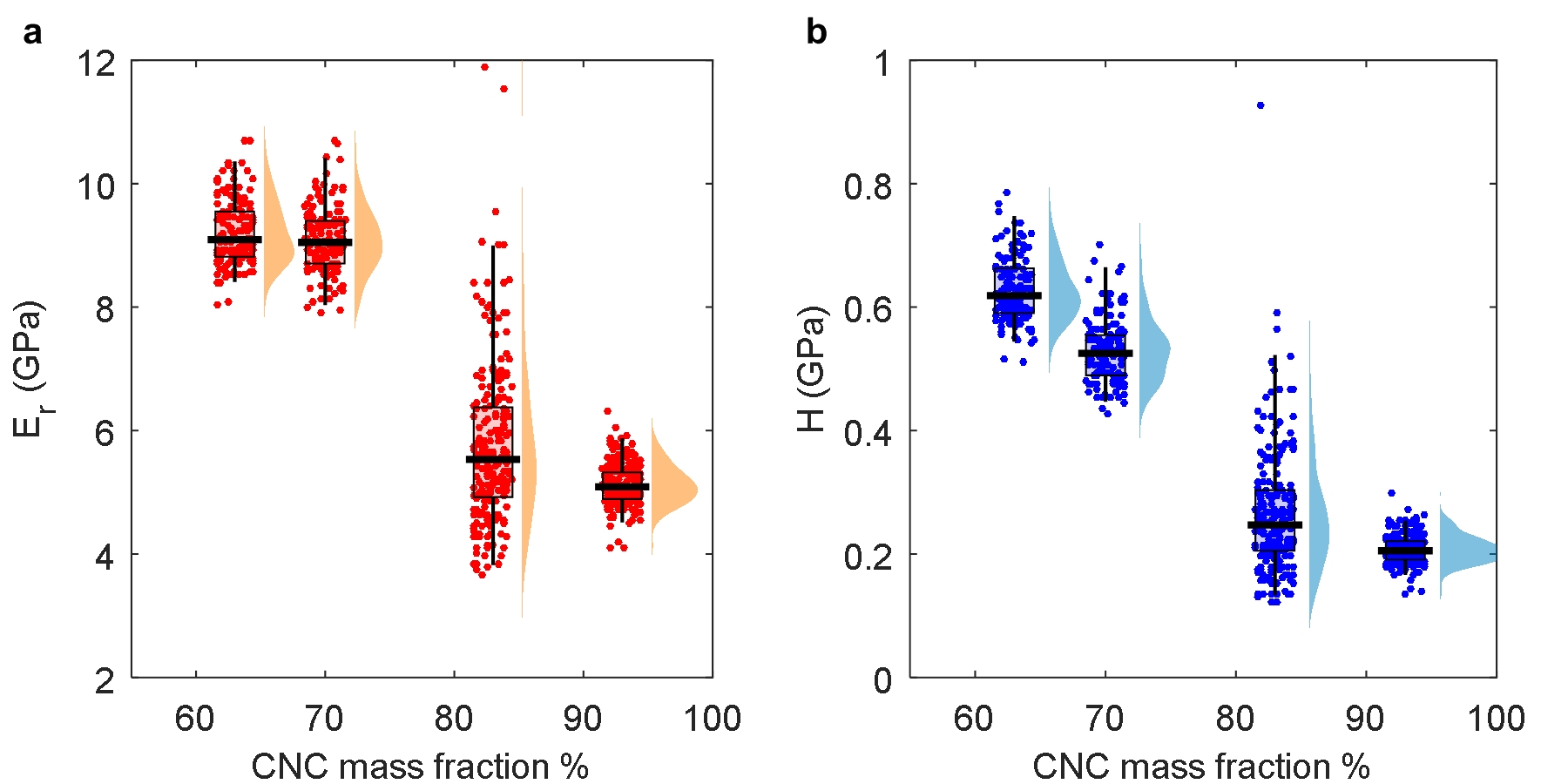}
	\caption[Effect of CNC mass fraction on modulus and hardness]{Variation in a) elastic modulus $E_{\rm r}$ and b) hardness $H$ of printed CNC-epoxide composites with respect to CNC mass fraction.}
	\label{printed_cnc_mass_fraction}
\end{figure*}

Figure~\ref{printed_cnc_mass_fraction} shows the elastic modulus and hardness of printed CNC composites with different CNC mass fractions, measured at a peak load of 10~mN. There is no improvement in modulus between a CNC mass fraction of 70\% and 63\%. Hence the CNC fraction of 63\%~wt. is chosen as the optimum for further testing. 

\section{Scratch test}
\begin{figure}
    \centering
	\includegraphics[width=0.8\columnwidth]{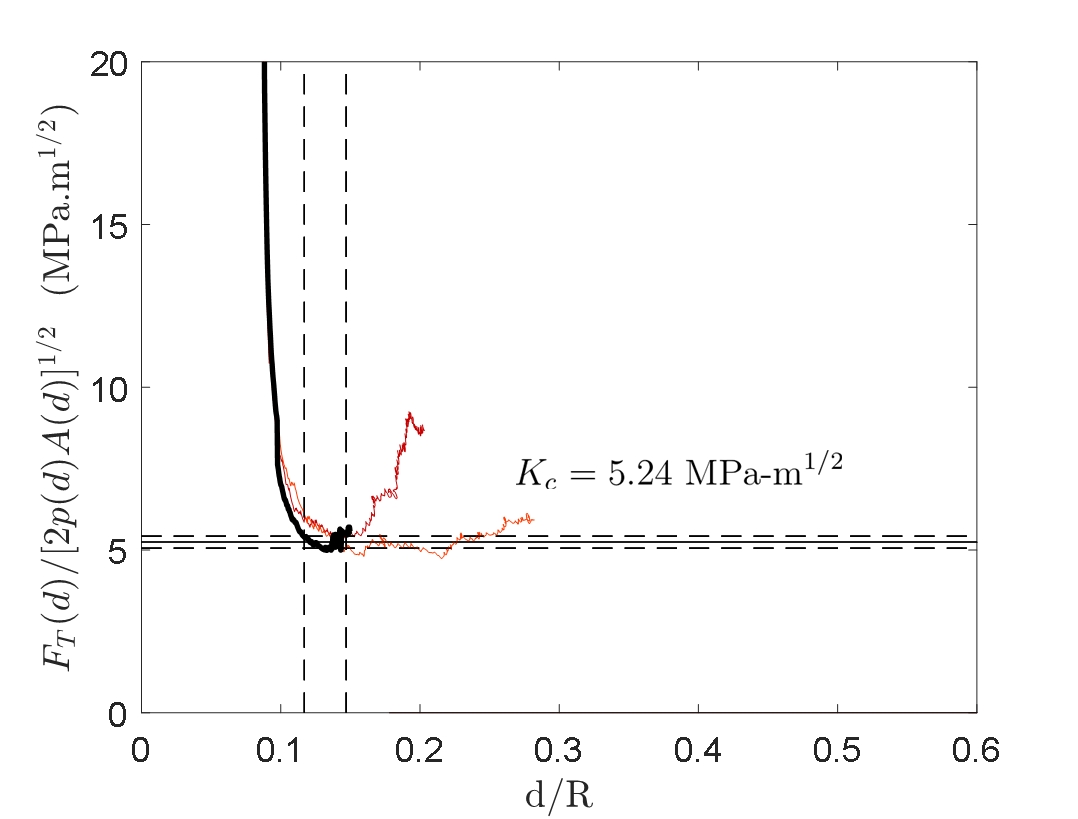}
	\caption[Scratch test of a printed composite]{Force versus normalized depth for two scratch tests on 63\% wt.~CNC/BADGE-ECH composite. Each scratch was performed at a velocity of 6~mm/min, with vertical force increasing from 30~mN to 30~N.}
	\label{CSA214_scratch_Kc}
\end{figure}

The fracture toughness $K_{\rm c}$ of the composites are determined from scratch tests performed with a Rockwell indenter connected to a Revetest tester (Anton Paar), following the method introduced in ref.\cite{Akono2011}. The vertical load on the indenter was increased linearly from 30~mN to 30~N over a scratch length of 3~mm. 
Fig.~\ref{CSA214_scratch_Kc} shows the normalized tangential force $F_T/\left(2pA\right)^{1/2}$ vs. $d/R$, where $d$ denotes the depth of the indenter and $R$ its radius (here $R=200~\mu$m), measured for a 3~mm long scratch test performed at a velocity of 6~mm/s, with a vertical force that increases from 30~mN to 30~N. For large enough depths, the ratio $F_T/\left(2pA\right)^{1/2}$ converges towards a constant value, which is reported as the rate-independent fracture toughness. Here, the CNC-epoxide composite with 63\% wt. CNCs has a fracture toughness $K_{\rm c}=(5.2\pm$0.2)~MPa.m$^{1/2}$

\section{Viscoplastic finite element model}

In materials with a time-dependent elastic or plastic response, the reduced modulus and hardness obtained by the Oliver-Pharr method, may not sufficiently describe the material.\cite{Menk2012} The CNC composites creep during the hold period of the nanoindentation profile, suggesting that a closer investigation of the viscoplastic properties is necessary. Furthermore, the force-displacement curves obtained from nanoindentation contain information about the onset of plastic deformation, not captured by the Oliver-Pharr model-based fit of the unloading curve.\cite{Fischer-Cripps2000} Here a finite element model is used to fit the force-depth data obtained from nanoindentation, thereby providing an estimate of the yield stress, Poisson's ratio, strain hardening and viscoplastic flow parameters. 

The model is based on an isothermal reduction of the Anand model for large-deformation, isotropic viscoplasticity.\cite{Narayan2018, Wang2001} The curve used for fitting is selected from the grid to be the one with $h_c$ closest to the mean value of $h_c$. The following flow equation gives the equivalent plastic strain rate,
\begin{equation}
    \dot{\overline{\epsilon^p}} = \dot{\epsilon_0}\left(\frac{\overline{\sigma}}{S}\right)^{1/m}
\end{equation}
where $\epsilon_0$ is the reference strain rate, $\overline{\sigma}$ is the magnitude of the deviatoric stress and $m$ is the rate-sensitivity parameter. The strain-hardening behavior is given by
\begin{equation}
    \dot{S} = H_0\left(1 - \frac{S}{S^*}\right)^{a}
\end{equation}
where $S$ is the Piola stress, $H_0$, $S^*$ and $a$ are strain-hardening parameters, and $S^*$ represents the saturation value of $S$.

\begin{figure}
    \centering
	\includegraphics[width=0.9\columnwidth]{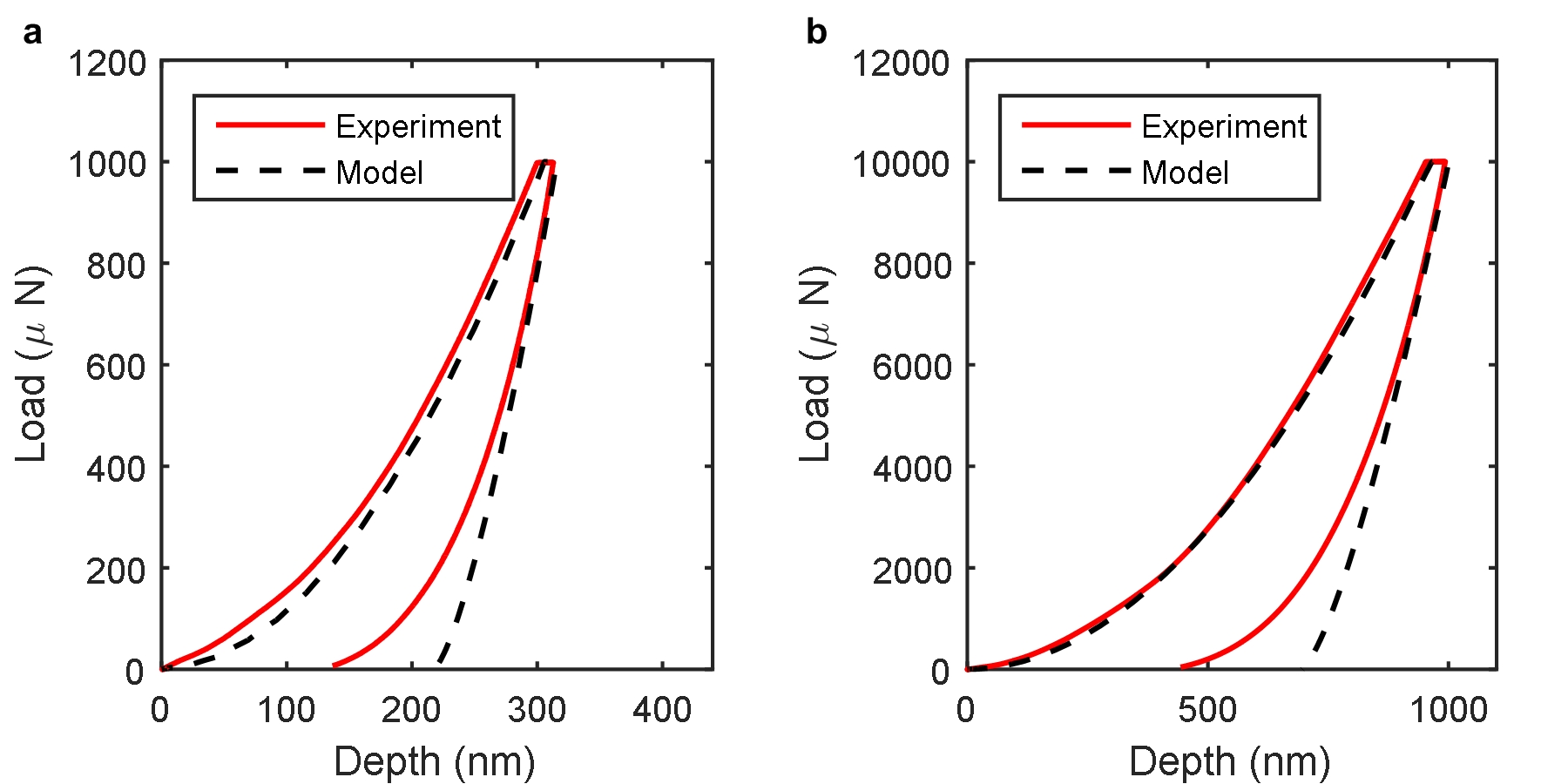}
	\caption[Viscoplastic model fits for the force-depth curves at peak loads of 1~mN and 10~mN]{Viscoplastic model fits for the force-depth curves at peak loads of: a) 1~mN and b) 10~mN}
	\label{FEM_1mN_10mN_fits}
\end{figure}

\begin{table*}[h!]
    \centering
    \begin{tabular}{ | m{2.5cm} | m{3cm}| m{3cm} |} 
    \hline 
    Parameter & $P_{max} = 1$~mN & $P_{max} = 10$~mN\\ 
    \hline \hline 
    Loading rate & 0.1~mN/s & 1~mN/s\\ 
    \hline
    $\dot{\epsilon_0}$ & 0.12 & 0.10\\ 
    \hline
    $m$ & 0.06 & 0.04\\ 
    \hline
    \end{tabular}
    \caption{Fit parameters describing rate-dependence for indents performed at peak loads of 1~mN and 10~mN with loading rates 0.1~mN/s and 1~mN/s respectively.}
    \label{FEM_fit_parameters}
\end{table*}

The force-depth curve corresponding to the mean plastic deformation, $h_c$, is selected from each distribution for fitting. Fig.~\ref{FEM_1mN_10mN_fits} shows the force-depth curves from experiment and the finite element model. The model and experimental data show good agreement for the loading segment and the upper 60\% of the unloading segment. The final plastic deformation is not accurately predicted due to difficulties in capturing the contact between the indenter and the sample surface. The Young's modulus and Poisson's ratio obtained from the fit are $E=8.5$~GPa and $\nu=0.2$ respectively. The yield strength from the fit, $\sigma_y=150$~MPa. These fit parameters are used for both values of $P_{max}$. However, the loading rates for each dataset are different since the indents were performed with the same loading profile, consisting of 10~s loading and unloading segments, separated by a 5~s hold. Thus, small changes are observed in the parameters that describe the rate-dependence of plastic deformation. Table~\ref{FEM_fit_parameters} lists the reference strain rate, and rate-sensitivity parameters for each of the fits. 

\section{Polarized microscopy}
\begin{figure*}
	\includegraphics[width=1.3\columnwidth]{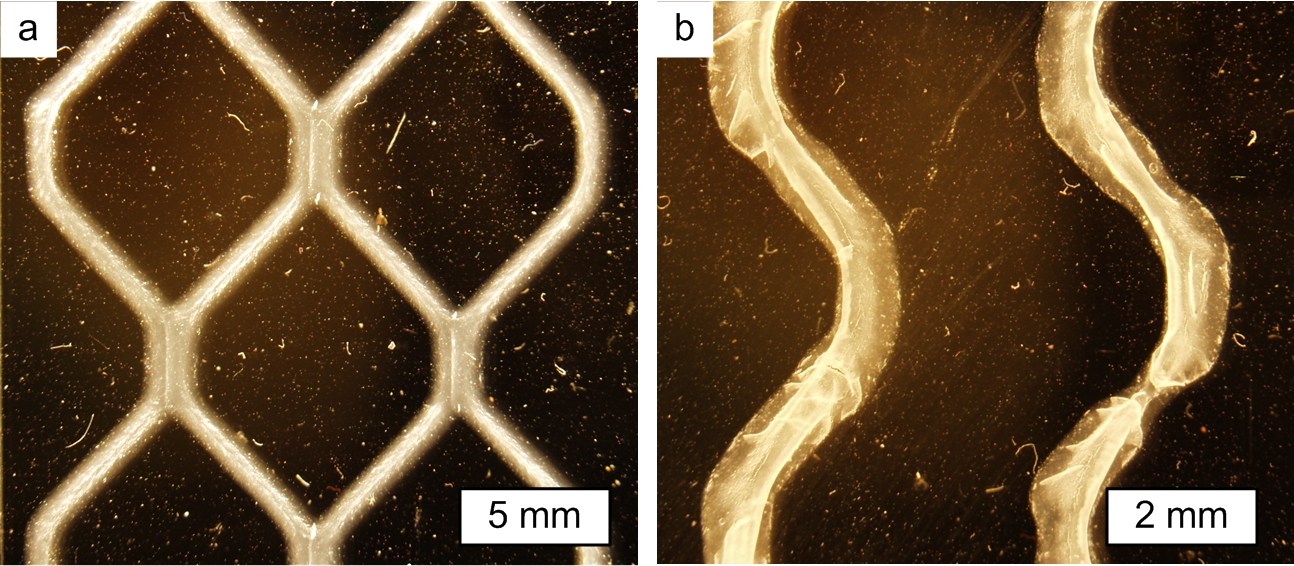}
	\caption{Transmission optical microscopy performed between crossed polarizers of cured printed traces with 63\% wt.~CNCs by mass: a) honeycomb pattern, b) curvilinear patterns.}
	\label{FigS2}
\end{figure*}

Hardened composites display an anisotropic microstructure as a result of the printing process. Images of 3D printed hardened composites (Fig.~\ref{FigS2}) were performed between crossed polarizers with a stereo-microscope (Olympus SZX16). The photos reveal a brighter inner section, with a duller outer section. The inner part corresponds to a region of the gel that is not fluidized during the printing phase and thus retains its alignment better. In the outer region, the hydrogen bonds are broken locally as the shear stress is the highest at the walls. Defects due to voids in the extruder are also visible. 

\section{Micromilling of CNC-polymer composite}
Bulk CNC artifacts were fabricated by micromilling (Roland SRM-20) a bulk composite of the cellulose nanocrystal composite. The composite was formed by molding the material into a 10~cm x 5~cm x 5~cm volume and thermally curing. Cutting parameters for a 0.0190" diameter square endmill were 9500 revolutions per minute and feed of 270~mm~min$^{-1}$. 

\end{document}